\documentclass[english,prb,twocolumn,showpacs]{revtex4}
\usepackage[T1]{fontenc}
\usepackage[latin1]{inputenc}

\makeatletter

\providecommand{\tabularnewline}{\\}

\usepackage{babel}
\makeatother
\begin{document}

\title{\emph{Ab initio} real-space Hartree-Fock and correlated approach
to optical dielectric constants of crystalline insulators }

\author{Priya Sony and Alok Shukla}

\affiliation{Physics Department, Indian Institute of Technology, Powai, Mumbai
400076, INDIA}

\begin{abstract}
In this paper we present an approach aimed at calculating the optical
dielectric constant ($\epsilon_{\infty}$) of crystalline insulators
both at the Hartree-Fock, and correlated levels. Our scheme employs
a real-space methodology, employing Wannier functions as the basic
building blocks. The scheme has been applied to compute $\epsilon_{\infty}$
for LiF, LiCl, MgO, and Li$_{2}$O. In all the cases, results of correlated
calculations, based upon systematic many-body corrections beyond Hartree-Fock,
exhibit excellent agreement with the experiments. To the best of our
knowledge, the calculations presented here are the first of their
kind for bulk solids.
\end{abstract}

\pacs{77.22.-d, 71.10.-w, 71.15.-m}

\maketitle

\section{Introduction}

\label{sec-intro}

It is of considerable interest to be able to compute various properties
of solids using an \emph{ab initio} methodology based upon their many-particle
wave functions\cite{fulde}. Such an approach allows one to initiate
the calculations at the Hartree-Fock level, and then systematically
improve them by including the electron-correlation effects to the
desired level of sophistication\cite{fulde}. In our recent work,
we have developed such a methodology for performing electronic structure
calculations on crystalline insulators\cite{shukla1,shukla2,shukla-jcp,shukla3,shukla4}.
The approach employs a real-space philosophy with Wannier functions
(WFs) used as the single-particle basis\cite{shukla1,shukla2,shukla-jcp}.
Electron-correlation effects are subsequently included by employing
the ``incremental scheme'' which is nothing but a Bethe-Goldstone-like
many-body hierarchy, consisting of virtual excitations into localized
unoccupied orbitals\cite{stoll}. Early applications of this approach
consisted of calculations of ground state geometries, cohesive energies,
and elastic properties of crystalline insulators at the Hartree-Fock
level\cite{shukla1,shukla2,shukla-jcp}, as well as at the many-body
level\cite{shukla3,shukla4}. Recently, we extended the approach to
perform both Hartree-Fock and many-body calculations of the Born effective
charges of insulators using a finite-field approach\cite{sony}. In
the present work we further extend our approach to compute a very
important quantity associated with dielectrics, namely the optical
dielectric constant $\epsilon_{\infty}$. Optical dielectric constant
is a quantity to which electron correlation effects can make significant
contributions. The fact that it is difficult to compute this quantity
accurately is obvious because it is only in the second-order perturbation
expansion of the total energy that the corrections due to the polarizability
of the system make their first appearance. In this paper we demonstrate
that our \emph{ab initio} Wannier-function-based methodology can also
be used to perform accurate calculations of $\epsilon_{\infty}$,
by reporting results on bulk LiF, LiCl, MgO, and Li$_{2}$O. The theoretical
results in all the cases are in very good agreement with the experiments.
Besides exploring the influence of electron-correlation effects on
$\epsilon_{\infty}$, we also examine the influence of the choice
of basis functions on this quantity. We conclude that the best computation
of electron-correlation effects is achieved only if the basis set
is of good quality. Additionally, wherever possible, we also compare
our results on $\epsilon_{\infty}$ to those obtained using the density-functional
theory (DFT) based approaches. 

Remainder of this paper is organized as follows. In section \ref{sec-theory}
we briefly discuss the theoretical aspects of our approach. Next,
in section \ref{sec-calc} we present and discuss the results of our
\emph{ab initio} calculations performed on various systems. Finally,
in section \ref{sec-conclusions} we present our conclusions.

\section{Theory}

\label{sec-theory}

Starting point of our approach is the expression of the electronic
contribution to the polarization per unit cell (dipole moment per
unit volume, ${\bf P}^{(e)}$) defined as the expectation value\cite{exp-r} 

\begin{equation}
{\bf P}^{(e)}=\frac{q_{e}}{N\Omega}\langle\Psi_{0}|{\bf R}_{e}|\Psi_{0}\rangle,\label{eq-pe}\end{equation}

where $\Omega$ is the volume of the unit cell, $q_{e}$ is the electronic
charge, $N\:(\rightarrow\infty)$, represents the total number of
unit cells in the crystal, ${\bf R}_{e}=\sum_{k=1}^{N_{e}}{\bf r_{k}}$
is the many-particle position operator for the $N_{e}$ electrons
of the crystal, and $|\Psi_{0}\rangle$ represents the correlated
ground-state wave-function of the infinite solid. From the polarization,
the dielectric susceptibility tensor (per unit cell) of the solid
($\chi_{ij}$) can be computed as

\begin{equation}
\chi_{ij}=\frac{\partial{\bf P}_{i}^{(e)}}{\partial{\bf E}_{j}}\label{eq-polarizability}\end{equation}

where, ${\bf E}$ is the external electric field, and $i$ and $j$
refer to the Cartesian components of the vector quantities involved.
For the cubic crystals considered here, only the diagonal components
of $\chi_{ij}$ are nonzero, which we denote as $\chi$, henceforth.
Finally, the dielectric constant of the crystal can be computed from
the well know expression $\epsilon=1+4\pi\chi$. By computing the
polarization of the crystal for atomic positions corresponding to
its ground state (i.e. no relaxation in the presence of applied electric
field), the dielectric constant obtained will be the optical dielectric
constant $\epsilon_{\infty}$ of the crystal. In the present work,
we avoid the tedious route of calculating first the correlated many-particle
wave function of the system ($|\Psi_{0}\rangle$), followed by the
expectation value of the dipole operator (cf. Eq. (\ref{eq-pe}))
and its subsequent derivative. Instead, we use the generalized Hellman-Feynman
theorem, and the finite-field approach of computing dipole expectation
values\cite{mcweeny}. Accordingly, we employ the modified Hamiltonian\begin{eqnarray}
H'({\bf E}) & = & H_{0}-q_{e}{\bf E}\cdot\sum_{k=1}^{N_{e}}{\bf r}_{k},\label{eq:ffield}\end{eqnarray}
where $H_{0}$ is the usual Born-Oppenheimer Hamiltonian for the solid,
and ${\bf E}$ is a user specified external electric field\cite{nunes}.
This leads to the modified Hartree-Fock equations in the Wannier representation
\begin{equation}
(T+U+2J-K+\lambda P_{{\cal E}}-q_{e}{\bf {\bf E}\cdot}{\bf r})|\alpha\rangle=\epsilon_{\alpha}|\alpha\rangle,\label{eq-hf}\end{equation}
where $T$ represents the kinetic-energy operator, $U$ is the electron-nucleus
potential energy, $J$ is the Coulomb term (or the direct term) of
the electron-electron repulsion, $K$ is the corresponding exchange
term, $P_{{\cal E}}$ is the projection operator constructed from
the WFs in the short-range environment ${\cal E}$ of the reference
unit cell, $\lambda$ is a large shift parameter, $|\alpha\rangle$
is one of the WFs in the reference unit cell, and $\epsilon_{\alpha}$
is the corresponding eigenvalue. The projection operator $P_{{\cal E}}$,
coupled with $\lambda$, plays the role of the localizing potential,
leading to self-consistent calculation of WFs localized in the reference
unit cell.\cite{shukla1,shukla2} The notations used here are consistent
with those in our original work aimed at direct determination of WFs
of crystalline insulators\cite{shukla1,shukla2}, and are explained
there in detail. The novelty of Eq. (\ref{eq-hf}), as compared to
the original Hartree-Fock equations\cite{shukla1,shukla2}, is the
presence of the term corresponding to the electric field. Thus the
WFs $|\alpha\rangle$ above, are obtained in the presence of an external
electric field, and therefore include the effects of dielectric polarization
as also enunciated by Nunes and Vanderbilt for tight-binding Hamiltonian\cite{nunes}.
One has to be careful in choosing the values of the electric field
${\bf E}$, because very large values can destroy the lower-bounds
of the spectrum leading to catastrophic results\cite{nunes}. In the
present work the maximum magnitude $|{\bf E}|=0.01$ a.u. was used,
and it did not cause any numerical problems. 

Once the Hartree-Fock equations for the system---with and without
the external electric field ${\bf E}$---have been solved, the correlation
effects are included by adopting the approach described in our earlier
papers.\cite{shukla3,shukla4,sony} For both types of many-body calculations
(i.e. with and without the electric field), the correlation energy
per unit cell is computed as per {}``incremental expansion''\cite{stoll},
\begin{eqnarray}
E_{corr} & = & \sum_{i}\epsilon_{i}+\frac{1}{2!}\sum_{i\neq j}\Delta\epsilon_{ij}+\nonumber \\
 &  & \frac{1}{3!}\sum_{i\neq j\neq k}\Delta\epsilon_{ijk}+\cdots\label{eq:inc}\end{eqnarray}
where $\epsilon_{i},\:\Delta\epsilon_{ij},\:\Delta\epsilon_{ijk},\ldots$
etc. are respectively the one-, two- and three-body$,\ldots$ correlation
increments obtained by considering simultaneous virtual excitations
from one, two, or three occupied WFs, and $i,\: j,\: k,\ldots$ label
the WFs involved\cite{shukla3}. The total correlated energy/cell
$E_{cell}$ can be simply obtained by adding $E_{corr}$ to the Hartree-Fock
energy $E_{HF}$ of the system, \emph{i.e.}, $E_{cell}({\bf E})=E_{HF}({\bf E})+E_{corr}({\bf E})$.
In the previous equation we have included the external electric field
${\bf E}$ explicitly in the parentheses to highlight the fact that
all the energies involved in the expression depend upon it. As in
our earlier works, the correlation approach used for computing various
increments was the full configuration-interaction (FCI) method, and
the incremental expansion was restricted to the two-body terms\cite{shukla3,shukla4,sony}.
For all the materials investigated, there are four valence WFs ($ns$
and three $np$ functions, $n$ being the principal quantum number
of the valence band) per unit cell. Therefore, the number of one-
and two-body increments for these materials is quite large.  However,
by using various point-group symmetries, we significantly reduced
the total number of increments actually computed. Finally, the susceptibility/cell
of the system concerned is obtained using the relation\begin{equation}
\chi_{ij}=-\frac{\partial^{2}E_{cell}}{\partial{\bf E}_{i}\partial{\bf E}_{j}}.\label{eq-chi}\end{equation}
The second derivative of the total energy/cell, with respect to the
external electric field needed in the equation above is computed numerically,
using the well-known central-difference formula.

\section{Calculations and Results}

\label{sec-calc}

Present calculations were performed using lobe-type Gaussian basis
functions employed in our earlier works as well\cite{shukla1,shukla2,shukla3,shukla4,shukla-jcp,sony}.
For MgO, we used the basis set used by McCarthy and Harrison\cite{mgo-basis}.
For Li$_{2}$O, LiCl, and MgO we went beyond the sp-type basis sets,
and augmented them with a single $d$-type function centered on the
anion, with the exponents $0.8$ for O in MgO, $0.65$ for O in Li$_{2}$O,
and $0.9$ for Cl in LiCl. The short range environment region ${\cal E}$
(cf. Eq.(\ref{eq-hf})) was taken to include up to third-nearest neighbors
of the reference unit cell. Therefore, the unit cell WFs for all the
materials studied were described using basis functions centered in
the cells as far as the third-nearest neighbors of the reference cell\cite{shukla1,shukla2}.
For all the compound we used the experimental values of the lattice
constants which were of 4.19 \AA$~$ for MgO, 4.573 \AA$~$ for Li$_{2}$O,
3.99 \AA$~$ for LiF, and 5.07 \AA$~$ for LiCl. For all the systems,
the fcc geometry was assumed. In MgO, LiF, and LiCl, the anion and
the cation locations in the primitive cell were taken to be $(0,0,0)$
and $(a/2,0,0)$, respectively, while in Li$_{2}$O, the anion was
located at $(0,0,0)$, and cations were located at positions $(\pm a/4,\pm a/4,\pm a/4)$,
where $a$ is the lattice constant. During the correlated calculations,
for LiF and Li$_{2}$O, the $1s$ WFs of both the cations and the
anions were treated as frozen core. In MgO, $1s$ WFs of both the
atoms, while $2s$, and $2p$ WFs located on the Mg were held frozen.
In case of LiCl, $1s$ WF of Li, and $1s,$ $2s$, and $2p$ WFs of
Cl were held frozen, while $3s$, and $3p$ chlorine WFs were correlated.
Thus, during the correlated calculations, the virtual excitations
were considered only from the valence WFs of the respective compounds.
For numerical calculations of the susceptibility (see Eq. (\ref{eq-chi})),
we used the central difference formula for the second derivative,
along with the values of the external electric field ${\bf E}$ ranging
from $\pm0.001$ a.u. to $\pm0.01$, in the $x$ direction. 

Results of our calculations are presented in tables \ref{tab-res}
and \ref{tab-cont}. Finally, in table \ref{tab-comp}, our best results
are compared with the experimental values, and also with those of
other authors. First, we examine Table \ref{tab-res} which presents
the calculated values of $\epsilon_{\infty}$ for LiF, Li$_{2}$O,
LiCl, and MgO obtained at the HF level, and at various levels of correlation
treatment, employing different basis sets. As mentioned earlier, we
have included correlation corrections using a real-space Wannier-function-based
approach developed earlier by us\cite{shukla3,shukla4,sony}. In the
present case, the included correlation contributions correspond to
one- and two-body increments confined to the WFs in the reference
cell. Thus two-body contributions corresponding to WFs in the nearest-neighboring
cells (and beyond) have not been included. Elsewhere, we will report
that the correlation contributions to $\epsilon_{\infty}$ arising
from two-body increments decrease rapidly with the increasing distances
between the WFs involved. Therefore, the inclusion of correlation
increments reported here will suffice for most cases. When we compare
the results of our best calculations to the experimental ones (cf.
table \ref{tab-comp}), we find that generally the agreement between
the two is excellent. The maximum disagreement is for MgO, with a
relative error of 6.8\%. For other three compounds the relative error
is well below 5\%. Next we analyze the results of these materials
one-by-one. Upon comparing the results of our Hartree-Fock values
of $\epsilon_{\infty}$ for LiF with the experimental one, we obtain
very good agreement between the two with the theoretical value being
slightly smaller than the experimental one. Next, when we include
the correlation effects, the theoretical value increases a bit, but
still stays in a excellent agreement with the experimental value.
For the case of MgO, we note that the results obtained with the sp
basis set are in poor agreement with the experiments, the disagreement
being $\approx$27\%. Upon including the single $d$-type exponent,
the improvement in the results begins at the HF level itself, with
the correlation contributions eventually bringing the results in close
agreement with the experiments. In table \ref{tab-res}, similar trends
are seen for the remaining compounds Li$_{2}$O and LiCl in that the
results are generally in poor agreement with the experiments if only
the sp basis sets are used, however, inclusion of $d$ exponent brings
them in excellent agreement with the experiments. Noteworthy point
is that it is only with the inclusion of $d$ exponents that correlation
contributions are significant, a point which is well-known in the
quantum-chemistry community in the context of polarizability calculations\cite{q-chem-pol}.
Now the question arises as to why excellent results were obtained
for LiF just with the sp basis set, while for MgO, Li$_{2}$O, and
LiCl inclusion of the $d$ exponents was essential. The fact that
for LiF, excellent results are obtained even without the use of d-type
functions, is a clear indication of the fact that in F$^{-}$ ion
valence electrons are extremely localized owing to the large electron
affinity of fluorine, and, therefore, can be described well with an
sp basis set. For Li$_{2}$O we attribute the large influence of $d$-type
functions to the extremely diffuse nature of the valence electrons
of O$^{--}$ ion. As a matter of fact, O$^{--}$ ion does not exist
in free form, and is stabilized only because of the crystal field.
Similarly, Cl$^{-}$ ion is comparatively more diffuse as compared
to the F$^{-}$ ion, thereby requiring the use of the $d$-type functions.
The other trend which is clearly visible from table \ref{tab-res}
is that when the calculations are performed with a proper basis set,
inclusion of the electron correlation effects brings the results in
much closer agreement with the experimental values, as compared to
the HF results. 

Next in table \ref{tab-cont} we examine the relative contributions
different correlation increments to the value of $\epsilon_{\infty}$
of these materials. Inspection of the table reveals that some very
clear trends are visible regarding various contributions. In all the
cases it is seen that the inclusion of the one-body increments increases
the value of the $\epsilon_{\infty}$ with respect to the HF value.
As far as the contribution of the two-body increments is concerned,
table \ref{tab-cont} reveals that for all the materials except for
LiF, this contribution is negative when the sp basis set is employed.
This, in our opinion, is a consequence of an inadequate basis set.
Once the spd basis sets are employed for these materials, the contribution
of this increment becomes significantly larger in magnitude, and positive
in sign, to bring the final values of $\epsilon_{\infty}$ in excellent
agreement with the experiments. 

Finally, in table \ref{tab-comp} we compare the results of our calculations
with those of other authors. As far as the DFT-based approaches are
concerned, one very powerful approach aimed at computing $\epsilon_{\infty}$
is the density-functional perturbation theory (DFPT) method\cite{dfpt}.
But, for the materials considered here, we were unable to locate any
calculations of $\epsilon_{\infty}$ based upon DFPT. However, in
the literature we found several results on $\epsilon_{\infty}$ based
on other DFT-based approaches which we discuss next. Umari and Pasquarello\cite{mgo-ref}
presented \emph{ab initio} molecular dynamics based calculations of
$\epsilon_{\infty}$ of MgO, performed in the presence of a finite
electric field. Mei \emph{et al.}\cite{lifcl-theory} \emph{}performed
DFT-based calculations of $\epsilon_{\infty}$ of LiF and LiCl, using
localized charge densities. For LiF and several other materials, Bernardini
and Fiorentini\cite{lif-theory} reported calculations of $\epsilon_{\infty}$
using a ``polarization approach'' in which they performed DFT-based
calculations on slabs of the material to calculate macroscopic polarization.
Then they obtained the values of $\epsilon_{\infty}$ by using a relation
between the interface charge, polarization difference and $\epsilon_{\infty}$.
To the best of our knowledge no prior calculations of $\epsilon_{\infty}$
of Li$_{2}$O exist. When we compare our results on MgO with those
of Umari and Pasquarello\cite{mgo-ref}, it is obvious that the two
results are in excellent agreement with each other. For LiF, our results
and those of Mei \emph{et al.}\cite{lifcl-theory} agree almost perfectly,
while the ones of Bernardini and Fiorentini\cite{lif-theory} overestimate
the experimental value. For LiCl results of Mei \emph{et al.} overestimate
$\epsilon_{\infty}$ somewhat as compared to the experimental value,
while our results are almost in perfect agreement with the experiments.

\section{Conclusions}

\label{sec-conclusions}

In conclusion, we have presented an \emph{ab initio} wave-function-based
approach aimed at computing the optical dielectric constant of insulators
both at the Hartree-Fock level, and the correlated level, employing
a real-space Wannier-function-based formalism. The approach was applied
to compute the $\epsilon_{\infty}$ of crystalline LiF, Li$_{2}$O,
LiCl, and MgO, and very good agreement with the experimental results
was obtained. The attractive aspect of the approach is that the correlation
effects are organized as per a Bethe-Goldstone-like many-body hierarchy,
which makes itself amenable to systematic expansion or truncation.
The influences of various terms in this correlation hierarchy on the
value of $\epsilon_{\infty}$ was examined for all the systems, which,
in our opinion, leads to a deeper understanding of the many-body effects.
Additionally, the influence of the choice of basis set on the computed
values of $\epsilon_{\infty}$ was also explored. The results presented
in this work suggest that this methodology holds promise, and its
applications to more complex systems such as perovskites will be pursued
in future.\\

\begin{acknowledgments}
We thank Department of Science and Technology (DST), Government of
India, for providing financial support for this work under grant no.
SP/S2/M-10/2000. \\

\end{acknowledgments}
\begin{table}[h]

\caption{Calculated values of $\epsilon_{\infty}$ for MgO, LiF, Li$_{2}$O,
and LiCl obtained at various levels of approximations and using different
basis sets. Column with the heading HF refers to results obtained
at the Hartree-Fock level. Headings {}``one-body/two-body'' refer
to results obtained after including the corrections due to {}``one-body/two-body''
excitations from Wannier functions of the unit cell, to the HF value. }

\begin{tabular}{|c|c|c|c|}
\hline 
&
\multicolumn{3}{c|}{$\epsilon_{\infty}$}\tabularnewline
\hline
\hline 
System&
HF&
one-Body&
two-Body\tabularnewline
\hline 
LiF (sp basis set)&
1.924&
2.029&
2.044\tabularnewline
\hline 
MgO (sp basis set) &
2.11&
2.15&
2.14\tabularnewline
\hline 
MgO (spd basis set) &
2.33&
2.42&
2.76\tabularnewline
\hline 
Li$_{2}$O(sp basis set)&
1.839&
1.873&
1.857\tabularnewline
\hline 
Li$_{2}$O(spd basis set)&
2.09&
2.33&
2.62\tabularnewline
\hline
LiCl (sp basis set)&
1.887&
1.929&
1.911\tabularnewline
\hline 
LiCl (spd basis set)&
2.096&
2.139&
2.833\tabularnewline
\hline
\end{tabular}

\label{tab-res}
\end{table}

\begin{table}[h]

\caption{Contribution of various types of electron-correlation effects to
$\epsilon_{\infty}$. Both the one-body and the two-body contributions
arise from virtual excitations from the WFs located within the reference
cell. Heading two-body (O) implies the corrections due to simultaneous
excitations from two distinct Wannier functions located on the anion
in the reference unit cell. }

\begin{tabular}{|c|c|c|}
\hline 
&
\multicolumn{2}{c|}{$\epsilon_{\infty}$}\tabularnewline
\hline
\hline 
System&
one-body&
two-Body(O)\tabularnewline
\hline 
LiF(sp basis set)&
0.105&
0.015\tabularnewline
\hline 
MgO (sp basis set)&
0.039&
-0.008\tabularnewline
\hline 
MgO (spd basis set)&
0.087&
0.338\tabularnewline
\hline 
Li$_{2}$O(sp basis set)&
0.034&
-0.016\tabularnewline
\hline 
Li$_{2}$O(spd basis set)&
0.239&
0.295\tabularnewline
\hline
LiCl (sp basis set)&
0.042&
-0.018\tabularnewline
\hline 
LiCl (spd basis set)&
0.043&
0.694\tabularnewline
\hline
\end{tabular}

\label{tab-cont}
\end{table}

\begin{table}[h]

\caption{Comparison of our best values of $\epsilon_{\infty}$ of various
materials with those of other authors, and experiments.}

\begin{tabular}{|c|c|c|c|}
\hline 
System&
This work&
Other works&
Experiment\tabularnewline
\hline
\hline 
MgO&
2.76&
2.78$^{a}$&
2.96$^{d}$\tabularnewline
\hline 
Li$_{2}$O&
2.62&
---&
2.68$^{e}$\tabularnewline
\hline 
LiF&
2.04&
2.03$^{b}$, 2.19$^{c}$&
1.96$^{f}$\tabularnewline
\hline 
LiCl&
2.83&
2.95$^{b}$&
2.79$^{f}$\tabularnewline
\hline
\end{tabular}

\label{tab-comp}

\noindent $^{a}$Reference \cite{mgo-ref}

\noindent $^{b}$Reference \cite{lifcl-theory}

\noindent $^{c}$Reference \cite{lif-theory} 

\noindent $^{d}$Reference \cite{exp-mgo}

\noindent $^{e}$Reference\cite{exp-li2o}

\noindent $^{f}$Reference\cite{exp-lifcl}
\end{table}

\end{document}